\documentstyle[aps,prl,multicol,epsf]{revtex}
\begin{document}
\draft
\title{ Crossover behavior for complex  order parameter  
in high-$T_c$ superconductors     }
\author{E. V.L. de Mello\cite{email} } 
\address{Departamento de F\'isica, Universidade Federal Fluminense, 
Av. Litor\^ania s/n, Niter\'oi, R.J., 24210-340, Brazil}
\date{\today}
\maketitle
\begin{abstract}
A number of recent experiments have suggested the presence
of either real or complex components in the gap symmetry of 
high-$T_c$  superconductors (HTSC).
In this paper we introduce a novel approach to  study  the competition of
such complex order parameter mixtures  by
varying the position of the two-body attractive potential
in a two dimensional extended Hubbard Hamiltonian. 
We show that this procedure explain a number of experimental
results and on the theoretical side, it may be related 
with certain HTSC microscopic models 
like the spin fluctuation theory.   Following current 
trends we concentrate on the study of $d_{x^2-y^2}$ order parameter 
with a component of the type $d_{xy}$ or a s-wave like
$s_{x^2+y^2}$ and  $s_{xy}$ symmetry.  
We show that the position of the optimal  s-component peak 
changes  with the position parameter  $b$ while the d-component 
occurs always in the optimally region around hole 
content $\rho \approx 0.39$. 
These studies may be useful  to interpret some 
experimental data  and  to explain 
why similar experiments yield different gap symmetries.
\end{abstract}
\begin{multicols}{2}

\narrowtext

\section{Introduction}
 In order to understand the fundamental       mechanism for high-$T_c$ 
superconductors (HTSC), many experiments have attempted to find out
the properties of the superconductors pair wave function. Notwithstanding  this
effort  the nature of the orbital order parameter symmetry
(or energy gap) has not yet been settled, despite 
increasing evidences toward a major $d_{x^2-y^2}$ symmetry as it explains
a number of different experimental results\cite{Cox,Harlingen}.
Furthermore the d-wave state has also theoretical support from weak 
and strong-coupling approaches\cite{Scalapino} and from Monte Carlo and
numerical studies on a two dimensional Hubbard model\cite{Dagotto}.
On the other hand several experiments have also suggested  
the presence of real or complex mixture of order parameters. For instance,
c-axis Josephson tunneling data between twinned YBCO and a s-wave superconductor
were interpreted  by a condensate containing a
mixture of $d_{x^2-y^2}$ and $s$ order 
parameter\cite{Dynes,Dynes2}. This type of order parameter symmetry was
argued to be a consequence of the orthorhombicity  of the YBCO crystal
structure\cite{Walker,Carbotte}. Recently, Klemm at al\cite{Klemm} have
studied the symmetry operation  of the crystal groups relevant to 
Hg1001, YBCO and BSCCO to analyze the possibility of different order
parameter symmetry combinations for these HTSC and their results agree
with the c-axis Josephson tunneling experiments.
Furthermore the
possibility of complex mixtures of different symmetries has also been
suggested\cite{SBL} as it would induce time reversal symmetry breaking in
connection with magnetic defects
or small fractions of  a flux quantum $\Phi_0=hc/2e$  observed 
in YBCO powders. Similarly a complex admixture
of $d_{x^2-y^2}$ and $s$ symmetry was also proposed\cite{Fogel} to
explain the ab-oriented YBCO/I/Cu tunnel junctions data\cite{Covington}.
They proposed an interesting and very simple physical picture: Andreev
scattering near the YBCO surface causes strong pair dissociation. The
resulting quasiparticles may then be paired again also by a subdominant
$s$-channel that is less sensitive to surface pair breaking then the
dominant $d_{x^2-y^2}$-channel. Calculations minimizing the free energy
show\cite{Fogel} that such complex mixture can coexist at low temperatures.
Using a tight binding model in a mean field treatment Honerkamp et al\cite{HWS}
have arrived at a similar conclusion that a surface phase transition towards
a time-reversal symmetry breaking surface state carrying an 
$s+$i$d_{x^2-y^2}$ state may appear (below $T_c$).
Recently\cite{Laughlin} a complex mixture of $d_{x^2-y^2}+id_{xy}$ was
also proposed in order to explain a phase transition characterized by
magnetic field plateaus or kinks in the 
thermal conductivity\cite{Krishana} of $Bi_2Sr_2CaCu_2O_8$ compounds.

Based on the above discussion,  we study in  this paper 
the competition of the $d_{x^2-y^2}$ and others subdominants complex
symmetries of the gap function.  We develop in these calculations 
a  novel approach 
based on a change in the position of the 
attractive potential $V$  from its usual nearest neighbor 
position in the extended Hubbard Hamiltonian.  As it is well known, the
on-site Coulomb correlations may explain the antiferromagnetism of the low
doping regime, the large magnetic fluctuations and the semiconductor
like properties of the metallic phase\cite{Anderson,MRRT,MRR}. Furthermore
the presence of a small attractive interaction leads to phase separation
and, on a square lattice, yields a superconducting phase with a 
$d_{x^2-y^2}$ symmetry\cite{Dagotto}. Thus, the Hubbard Hamiltonian with
such phenomenological attraction
is a natural candidate to deal with the HTSC. The idea to change 
the position of the attractive potential in the extended Hubbard
Hamiltonian  
was originally motivated by the experimental fact that different compounds
have different $T_c\times n$  phase diagram and the optimal doping values
may vary according the compound. This fact can
be physically interpreted due to a possible  change in the range of the 
attractive interaction that leads to pair formation \cite{Mello1,Mello2}; 
for compounds with short range correlations the
carrier density must be larger than those compounds with a longer ranged
interaction in order to undergo a superconducting phase transition. 
The same type of physical argument  can be used to interpret
the measured different values of the coherence length $\xi$ for 
different family of  compounds since a short range interaction requires  
larger densities than a long range one in order to produce the coherence 
motion that leads to superconductivity. Furthermore, 
the typical  parameter which characterizes the strength of the
attractive interaction\cite{Landau}, 
$\chi  = m_ea^2V/2\hbar^2$,  when estimated for most of  HTSC, 
varies between 10-3000 and such high values indicate that  the size
of the bound states (or Cooper pairs) are indeed related to the
size of the minimum of the attractive potential. 

Another interesting aspect from the theoretical point of view is that 
this parametric change of the attractive potential can be related with 
the order parameter expansion introduced by the 
spin-fluctuation theory\cite{Pines}. In this 
approach  it has been proposed a $d_{x^2-y^2}$-wave gap of the 
form $\Delta_0(\vec k)(cosk_xa-cosk_ya)$ where $\Delta_0
(\vec k)$ is expanded in powers of $(cosk_xa+cosk_ya)$. Taking a 
close look at this expansion, one can easily verify that it contains 
terms like $(cosk_xa-cosk_ya)(cosk_xa+cosk_ya)$ which are
proportional to  $(cos(2k_xa)-cos(2k_ya))$ and what can be seen as 
a type of d-wave
gap symmetry that arises from a potential 
like $V(\vec k)=V_0(cos(2k_xa)+cos(2k_ya))$. By the same token, we
can find  terms proportional to  $(cos(3k_xa)-cos(3k_ya))$
which can be originated by a potential  $V(\vec k)=V_0(cos(3k_xa)+cos(3k_ya))$
and so on. There will also be crossed terms proportional 
to $(cosk_xa\times cosk_ya)$ that may be
associated to a next-nearest potential.  Thus, we see that the gap    
expression proposed by the spin-fluctuation theory contains several terms
and some may be associated to a potential of the form 
$V(\vec k)=V_0(cos(bk_xa)+cos(bk_ya))$, where $b=1,2,3,...$ is  a parameter
related  to the position of the attractive potential 
($b=1$ leads to  the usual nearest-neighbor expression).  
Consequently we have studied in this paper the phase diagram of
complex admixtures of a  d-wave like order parameter with
some minor complex component with this parameterized change of the
potential position, that is, with different values of $b$. 
These calculations are made with the extended
Hubbard model on a square lattice of side $a$, which describes
a tight binding model with the strong correlations taken into account
and is defined by
\begin{equation}
H=-\sum_{\langle ij \rangle , \sigma}t_{ij}(c^{\dagger}_{i \sigma} c_{j \sigma}
 + h.c.)+U \sum_{i} n_{i \downarrow}n_{i \uparrow}-V \sum_{\langle ij
 \rangle}n_{i}n_{j} .
\label{Hamiltonian}
\end{equation}
where $t_{ij}$ is the  hopping integral between sites $i$ and $j$ and
are estimated by comparison with either band structure calculations
or Fermi surface measurements, U is the on-site correlated
repulsion and V is an attractive phenomenological interaction
which in principle can be due to spin fluctuations, as mentioned above
or any other type of mechanism. 
In order to obtain quantitative results, we have used  the 
tight-binding dispersion relation derived from Fermi surface measurements
by Norman et al\cite{Ding} for $Bi_2Sr_2CaCu_2O_8$ but this choice is
not  a fundamental requirement, another studies have used different 
dispersion\cite{Angilella}  with small quantitative changes.

\section{The Method} 
We used a BSC mean-field
approximation\cite{de Gennes,Legget} to study the superconducting phase
associated with the Hamiltonian of Eq.\ (\ref{Hamiltonian})
described above.  In this model the superconducting
state is characterized by a gap order parameter, which at a finite
temperature T satisfies the BCS self-consistent equation

\begin{equation}
\Delta_{\vec k}=-\sum_{\vec l}V_{\vec k\vec l}F_{\vec l}
\label{gapuv}
\end{equation}
where $V_{\vec k\vec l}$ is
the Fourier transform of the potential of Eq.\ (\ref{Hamiltonian}) and

\begin{equation}
F_{\vec k}=\frac{\Delta_{\vec k}}{2E_k}tanh\frac{E_k}{2k_BT} \: , \: E_k=\sqrt{
(\varepsilon_k-\mu)^2+\Delta_k^2}
\label{gapF}
\end{equation}
where $\varepsilon_k$ is the dispersion relation taken from Ref.\cite{Ding}
and $\mu$ is the chemical potential.
Following along the lines of Ref.\cite{Angilella} , one can
easily show that $V_{\vec k\vec l}$  
may be written in a  "separable" form,
$V_{\vec k\vec l}=U-2V\left(\cos(k_xa)\cos(l_xa)+\cos(k_ya)\cos(l_ya)\right)$
which leads to the usual d and s-wave.
This procedure can easily  be generalized for the case of $b=2,3...$ 
discussed above and  will be developed below in order to study
the superconducting transition with respect to the position
of the attractive potential. For a two-component order 
parameter with a separable form, the corresponding gap equations 
may be written as\cite{Hng1}
\begin{equation}
V_{\vec k \vec l}=\sum_{j=1}^2V_jf^j_{\vec k}f^j_{\vec l} \; , \; 
\Delta_{\vec k}= \sum_{j=1}^2\Delta_jf^j_{\vec k}
\label{gapV}
\end{equation}
Here, since we want to study the competition between the
$d_{x^2-y^2}$ and others symmetries, we have always 
taken $V_1=V_{d_{x^2-y^2}}$, 
$f^1_{\vec k}=(cos(bk_xa)-cos(bk_ya))$ and $\Delta_1=\Delta_{d_{x^2-y^2}}$ and
the complex component can either 
be $V_2=V_{d_{xy}},V_{s_{x^2+y^2}}$ or $V_{s_{xy}}$
with $f^2_{\vec k}=2sin(bk_xa)sin(bk_ya),(cos(bk_xa)+cos(bk_ya))$ or 
$2cos(bk_xa)cos(bk_ya)$ and the gap functions are $\Delta_1=i\Delta_{d_{xy}},
\Delta_{s_{x^2+y^2}}$ or $\Delta_{s_{xy}}$ respectively.
Now, separating the real and
imaginary parts and combining Eq.\ref{gapF} and Eq.\ref{gapV} we obtain
the following  two gap equations
\begin{equation}
\Delta_j= \sum_kV_j\frac{\Delta_jf^{j^2}_{\vec k}}{2E_k}tanh\left(
\frac{E_k}{2k_BT}\right) \: , \; j=1,2
\label{gapFf}
\end{equation}
These two equations has to be solved with  the density of 
carriers $\rho $ equation
which is given by\cite{Legget}
\begin{equation}
\rho (\mu,T)=\frac{1}{2}\sum_k\left(1-\frac{\varepsilon_k}{E_k}
tanh\frac{E_k}{2k_BT}\right)
\label{dens}
\end{equation}
 Thus we solve self-consistently the above 3 equations (Eq.\ref{gapFf}
and Eq.\ref{dens}) in order to study the
phase diagram of a complex admixture of the order parameter. 
These equations are general in the sense that they can be applied
to any type of two component symmetry mixtures.
The results will be discussed below.

\section{Results and Discussions}
\begin{figure}[!h]
\center
\epsfxsize=3.6truein
\centerline{\epsffile{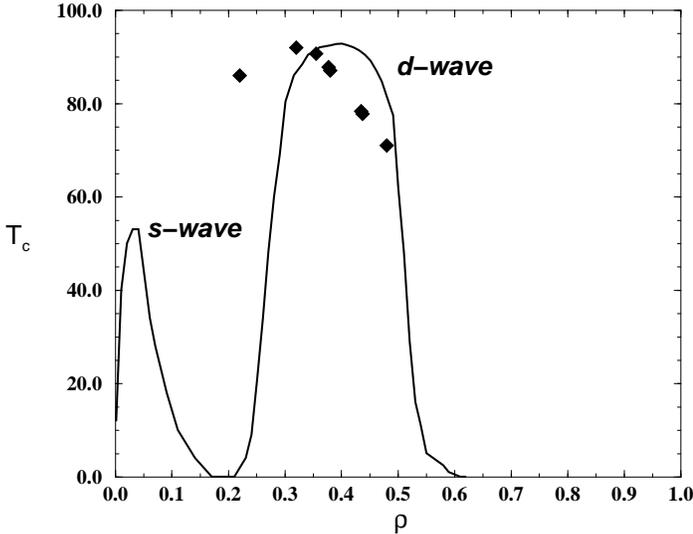}}
\caption{ The critical temperatures for the s and d-channels for
$b=1$. The continuous
line is the calculations described in the paper and the diamonds are the
experimental points from Ref. 28}
\label{fig:Tc}
\end{figure}
We start with the analysis for the complex mixing with
the two  most relevant\cite{Cox} symmetries, The $d_{x^2-y^2}$ 
and the $s_{x^2+y^2}$ which we write as  $d_{x^2-y^2}+is_{x^2+y^2}$.  
We have solved the above self consistent equations with fixed
$V_d/t=0.11$ and $V_{s_{x^2-y^2}}/t=0.18$, which these values were chosen in
order to obtain reasonable values for the critical temperatures $T_c$  
in the range of parameters studied. In this way we obtain a maximum
$T_c\approx 94$K for b=1  and by the  dominant $d_{x^2-y^2}$ symmetry near
the optimum doping. In Fig.1 we show the results for the values
of $T_c \times \rho$  in order to compare with the experimental results
of Allgeier et al\cite{Allgeier} on $Bi_2Sr2CaCu_2O_{8+\delta}$ (Bi2212). 
Notice that their maximum $T_c$
occur at $\rho \approx 0.32$ while our calculations yield a maximum $T_c$ 
at $\rho \approx 0.39$ and this discrepancy is probably due the
dispersion relation that we used since a similar BCS mean field
calculation using the Hubbard Hamiltonian (for b=1)
gives a maximum which agrees with the experimental result\cite{Angilella}.

We should point out that the comparison with a HTSC experimental phase diagram,
as in Fig.1, has only meaning if the material has a BCS type behavior, that is,
when the appearance of the superconducting gap and phase coherence occur 
simultaneously at the onset of superconductivity. On the 
other hand, angle-resolved photoemission spectroscopy 
(ARPES) on underdoped Bi2212 has revealed the presence of a   "pseudogap"
above $T_c$\cite{Marshall,Loeser,Ding} which has also been confirmed by recently
Fermi surface ARPES measurements\cite{Norman}. This unusual behavior may be an
indication of non-BCS behavior which $T_c$ is controlled by the doping level,
at underdoped regime, rather then by phase coherence with a single particle
gap $\Delta_{\vec k}$\cite{Emery}. Nevertheless this is not a completely settled
scenario which only more experiments will give the correct interpretation
specially because the pseudogap has not yet been seen in several cuprate
superconductors\cite{Coleman}. 
\begin{figure}[!h]
\epsfxsize=3.6truein
\centerline{\epsffile{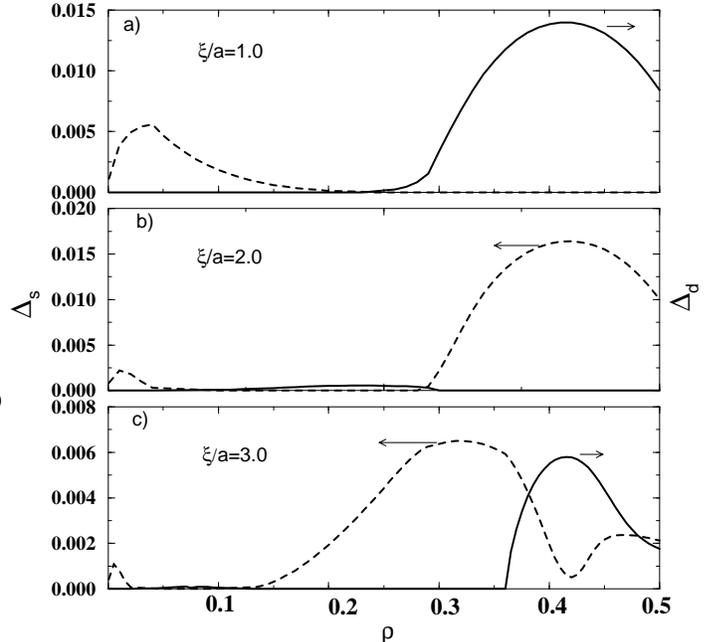}}
\caption{Results for the zero temperature $\Delta_{d(x^2-y^2)}$ (continuous lines) and 
$\Delta_{s(x^2+y^2)}$ (dashed lines)  in units of $t=0.26 eV$ as
function of the hole density $\rho$ for $b$=1(a),2(b) and 3(a). $\rho=0$
is the half-filling density. }
\label{fig:ds}
\end{figure}
In the Fig.2 we show the results for the zero temperature gap 
amplitudes $\Delta_{d_{x^2-y^2}}$ (continuous line) and  $\Delta_{s_{x^2+y^2}}$
(dashed line) as function
of the  density $\rho$ for three values of the parameter $b$, namely
$b=1,2$ and $3$. Where there is  superconductivity region, the values of $\Delta$  have 
their maxima at zero temperature and vanishes at $T_c$   in agreement with
the BCS method. In fact, the self-consistent
calculations yields the temperature, the density $\rho$ and the
value of the gap as one solves Eqs. 5 and 6 simultaneously. 
In Fig.3 we show how the gap changes with the temperatue. The zero temperature
gap value is proportional to $T_c$ and the constant of proportionality
differs  slightly  from one channel to the other.  We get   
$\Delta(0)/k_BT_c \approx 2.6$ in the optimally region.
\begin{figure}[!h]
\center
\epsfxsize=3.6truein
\centerline{\epsffile{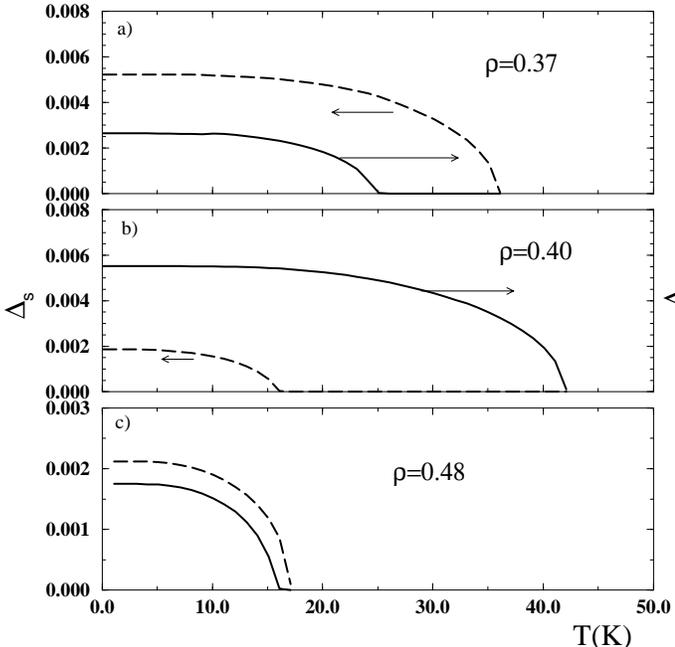}}
\caption{The display of the doubled crossover. As the density increases
we see the solutions change from dominant s-channel in a), to a dominant
d-channel as in b) and again a dominant s-channel in c). The temperature
T is in Kelvins.}
\label{fig:sdcross}
\end{figure}
There are very interesting  differences from the $s$ and $d$ channels
which arises as we vary the parameter $b$. For $b=2$, the values 
of $\Delta_{d_{x^2-y^2}}$ are basically
suppressed while for $b=3$ its its strength is decreased by about half of
its value for $b=1$ (which agrees with the calculations of 
Ref.\cite{Angilella}) and  does not move appreciably from 
the optimum doping value of  hole content $\rho=0.39$ 
($\rho$ is the fraction of holes per Cu atom in the $CuO_2$ sheet).
This result is a little higher then the experimental result\cite{Allgeier}
which yields $\rho=0.32$ for the Bi optimum doping value. The maximum
$T_c$ is related with the envelope of the curve  shown
in Fig.2.  On the other hand, for the 
extended $s$, $\Delta_{s_{x^2+y^2}}$,
the change in $b$ causes the value of the corresponding optimum doping density
to change continuously; it is at $\rho \approx 0.41$ for $b=2$ and and about
$\rho=0.32$ for $b=3$ as it is seen in Figs.2a, 2b and 2c. At $b=1$ (Fig.2a),
the s-wave has the optimum doping at $\rho \approx 0.05$, for $b=2$ the
maximum $T_c$ occurs at $\rho \approx 0.40$ and it dominates  over $\Delta_d$
which is  almost entirely suppressed. For
$b=3$ the competition between the two channels becomes very 
interesting; the maximum of $\Delta_s$ has moved to $\rho \approx 0.33$ while the
maximum of $\Delta_d$ remains  fixed at $\rho \approx 0.42$ which gives
rise to a completely novel {\it doubled crossover phenomena}; 
up to $\rho \approx 0.35$
the self-consistent equations yields pure extended s-wave solutions. Between
$\rho  \approx 0.35-0.37$ the mixing of order parameters starts at low
temperatures but with the
s-channel dominating over the d-channel as its  critical temperature
$T_c$ is larger as shown in Fig.2c and at Fig.3a for  $\rho=0.37$.

At $\rho \approx 0.38$  a crossover arises and the d-channel solutions 
dominates over  the s-channel which becomes the minor component as
shown in Fig.2c and Fig.3b. There is always mixing at low temperatures 
(the continuous curve of the minor s-wave never vanishes completely)
but the d-channel is the only one at $T_c$.
At  $\rho \approx 0.48$ there is a new crossover and 
the dominant solution changes  again as the s-wave
becomes the dominant solution and the only one at $T_c$ and the 
d-wave appears  again at low temperatures as the minor components as
displayed in Fig.2c and Fig.3c. Notice that the appearance of this
subdominant component is exactly as the mechanism proposed to
explain the ab-oriented YBCO/I/Cu tunnel junction data\cite{Covington}
discussed in the introduction.
\begin{figure}[!h]
\epsfxsize=3.6truein
\centerline{\epsffile{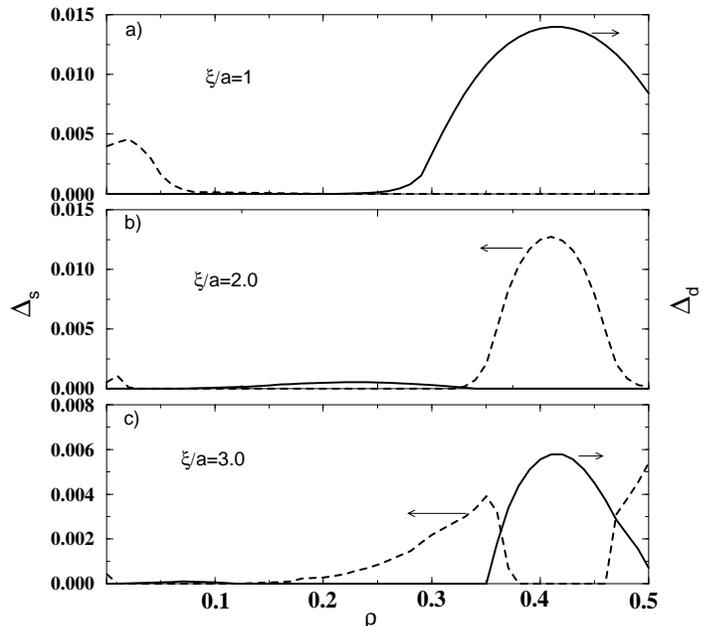}}
\caption{Calculations for the $\Delta_{d(x^2-y^2)}$ (continuous lines) 
and $\Delta_{s(xy)}$ (dashed lines)  in units of $t=0.26 eV$ as
function of the hole density $\rho $ for $b$=1(a),2(b) and 3(a).}
\label{fig:dsxy}
\end{figure}
Next we have studied the  complex gap function of  symmetry 
$d_{x^2-y^2}+is_{xy}$. The results are quite similar to those above; for
$b=1$ the d-state dominates and has its optimum doping at $\rho=0.40$
as shown in Fig.4a. For $b=2$ the d-channel is again suppressed
and the $s_{xy}$-state is the only one and
its maximum doping is also at $\rho \approx 0.83$ as displayed in Fig.3b. For
$b=3$ there is again the very interesting {\it doubled crossover} between these
two competing symmetries.  For $0.2<\rho <0.36$ there is only s-wave 
instability but at $\rho \approx 0.36$ a small d-wave channel develops at
low temperatures. Between $\rho \approx 0.38$ and $\rho \approx 0.45$ we find
only pure $d_{x^2-y^2}$ solutions as opposed to the case shown above in
Fig.2c where a low temperature mixing was always present. For $\rho >0.45$
the $s_{xy}$-channel starts to develop, initially as a minor component
(see Fig.5b) and above  $\rho=0.47$  a new crossover appears as 
it becomes the dominant channel and the only channel at $T_c$.
Thus we verified that the states $d_{x^2-y^2}+is_xy$ has also the {\it doubled
crossover} and the dominant regions are similar to those for 
$d_{x^2-y^2}+is_{x^2+y^2}$ however, there is a new feature with this type
of admixture: we can see from Fig.4c that, for $b=3$, one kind of gap symmetry
influences the other. One can easily see the abrupt change in $\Delta_{s_{xy}}$
near $\rho \approx 0.33$ and the change in slope for both channels near 
$\rho \approx 0.48$. The new phenomenon, that is, the
slightly suppression on $\Delta(T)$ of the dominant channel can also be 
seen on Figs.5a, 5b and 5c. This is a new feature  not seen in Figs.3c, and this
suppression is the cause of the pure d-wave solutions at the optimal region
which did not occur for previous studied  extended s-wave. This interesting 
feature may be similar to what happens in Andreev scattering near surfaces 
as discussed in the introduction.
\begin{figure}[!h]
\center
\epsfxsize=3.6truein
\centerline{\epsffile{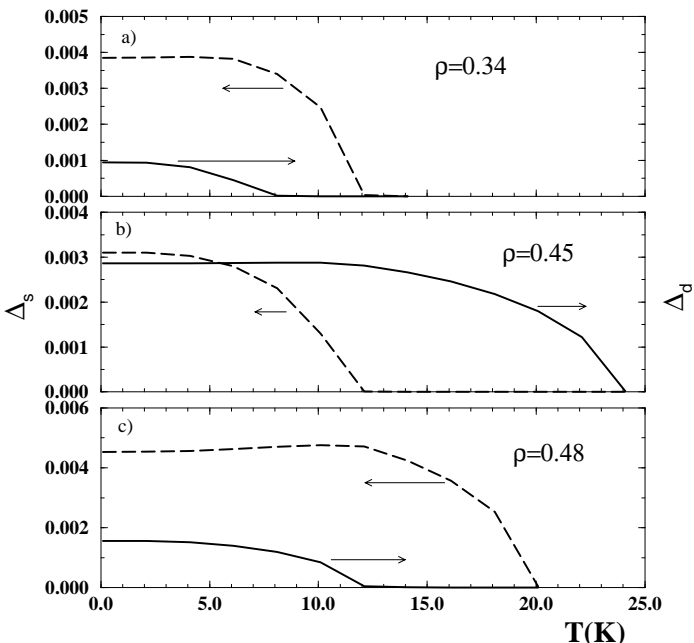}}
\caption{The display of the doubled crossover. As the density increases
we see the solutions change from dominant s-channel in a), to a dominant
d-channel as in b) and again a dominant s-channel in c). The temperature
T is in Kelvins.}
\label{fig:sdxycross}
\end{figure}
Lastly we have also studied the complex order parameter
mixing $d_{x^2-y^2}+id_{xy}$ for the
three values of b, however we have not found even a single crossover. For
$b=1$, the $d_{x^2-y^2}$ channel dominates and has also its optimum
value at $\rho=0.40$ and the $d_{xy}$ appears only at  low temperatures. 
For $b=2$ and $3$ the $d_{xy}$ dominates completely
around the same optimal region and the other channel gives only a very small
component at low temperatures. Thus we do not observe any crossover for
$d_{x^2-y^2}+id_{xy}$ complex mixture, simply the interchange 
of dominant symmetry as we change $b$.
\section{Conclusions}
The complex mixtures in superconducting order parameter were studied here as 
function of hole density  in the context of the BCS mean-field
calculations for the extended Hubbard model. 
We have revealed very interesting properties that may be
relevant for HTSC; if the interaction is sufficiently short ranged that
can be mapped by  a nearest neighbor potential, at low doping, the system
is described by either {\it pure} $d_{x^2-y^2}$ or 
$s_{x^2+y^2}$  order parameter. If the range of the potential is larger 
the system changes its order parameter
symmetry (for $b=2$) or may admit a mixed state with both $s$ and $d$
components present (for $b=3$). In this last case the dominant component
depends strongly on the density or doping level $\rho$ and may change
abruptly the dominant symmetry at $T_c$. Since in most HTSC it is
difficult to determine $\rho$ accurately\cite{Tallon} and as our calculations
demonstrate, a  small change in $\rho$ may be accompanied by a 
crossover of the dominant component. If we assume an expansion with 
$b=1,2,3...$  like that of the spin fluctuation theory of Ref.\cite{Pines},
then one must look at their weight
to see which one is more important and concomitantly, which symmetry of
the order parameter is the dominant one. This present work  provides 
a hint  to understand
why different order parameter symmetries appear in different experiments
performed on the same type of compounds. 

\section{Acknowledgments}
We  acknowledge  partial
financial support from the Brazilian agencies Capes and  CNPq and the
hospitality of the CNRS/CRTBT  at Grenoble, France.

\end{multicols}


\begin{references}
\bibitem[*]{email}  evandro@if.uff.br
\bibitem{Cox} D. L. Cox and M. B. Maple, Phys. Today, 32, fev. of 1995.
\bibitem{Harlingen} D. J. Van Harlingen, Rev. Mod. Phys, {\bf 67}, 515 (1995).
\bibitem{Scalapino} D. J. Scalapino, Phys. Rept. {\bf 250}, 329 (1995).
\bibitem{Dagotto} E. Dagotto, J. Riera, Y. C. Chen, A. Moreo, A. Nazarenko,
F. Alcaraz and F. Ortoloni, Phys. Rev. {\bf B49}, 3548 (1994).
\bibitem{Dynes} G.G. Sun, D. A. Gajewski, M. B. Maple and R. C. Dynes, Phys.
Rev. Lett. {\bf 72}, 2267, (1994).
\bibitem{Dynes2}K. A. Kouznestsov, G.G. Sun, B. Chen, A. S. Katz
S. R. Bahcall, John Clarke, R. C. Dynes, D. A. Gajewski, S. H. Han,
M. B. Maple , H. Giapintzakis, J. T. Kim and D. M. Ginsberg, Phys.
Rev. Lett. {\bf 79}, 3050, (1997).
\bibitem{Walker} M. B. Walker, Phys. Rev. {\bf B53}, 5835, (1996).
\bibitem{Carbotte} C. O'Donovan, D. Branch, J. P. Carbotte and J. S. Preston,
Phys. Rev. {\bf B51}, 6588, (1995).
\bibitem{Klemm} R. A. Klem, C. T. Rieck and K. Scharnberg, cond-mat 9811303,
submitted to Phys. Rev. B.
\bibitem{SBL} M. Sigrist, D. B. Bailey,  and R. B. Laughlin, Phys. Rev. Lett.
{\bf 74}, 3249, (1995).
\bibitem{Fogel} M. Fogelstr\"om, D. Rainier, and J. A. Sauls, Phys.
Rev. Lett.{\bf 79}, 281 (1997).
\bibitem{Covington} M. Covington, M. Aprili, E. Paraonu, L.H. Greene, 
F.Yu, J. Zhu and C. A. Mirkin, Phys. Rev. Lett.{\bf 79}, 277 (1997).
\bibitem{HWS} C. Honerkamp, K. Wakabyashi and M. Sigrist, 
cond-mat 9902026.
\bibitem{Laughlin}  R. B. Laughlin, Phys. Rev. Lett.  {\bf 80}, 5188, (1998).
\bibitem{Krishana} K. Krishana, N. P. Ong, Q. Li, G. D. Gu, and N. Koshizuka,
Science, {\bf 277}, 83, (1997).
\bibitem{Anderson} P. W. Anderson Science {\bf 235} 1196 (1987).
\bibitem{MRRT} R. Micnas, J. Ranninger, S. Robaszkiewicz and S. Tabor,
Phys. Rev. {\bf B37} 9410 (1988).
\bibitem{MRR} R. Micnas, J. Ranninger and S. Robaszkiewicz, Rev. Mod. Phys.,
{\bf 62}, 113 (1990).
\bibitem{Mello1} E.V.L.de Mello, Physica {\bf C259}, 109(1996) and 
Chec. J. Phys.{\bf 46}, 945 (1996).
\bibitem{Mello2} E.V.L. de Mello and C.Acha, Phys. Rev {\bf B56}, 466 (1997)
and Physica {\bf C 282-287}, 1819, (1997).
\bibitem{Landau} "M\'ecanique Quantique", L. Landau and I. Lifshitz,
(Ed. Mir, Moscow 1966), page 190.
\bibitem{Pines} P. Monthoux, A. V. Balatsky and D. Pines, Phys. Rev. Lett.
{\bf 67}, 3448 (1991).
\bibitem{Ding} M. R. Norman, M. Randeria, H. Ding and J. C. Campuzano, Phys.
Rev. {\bf B52}, 615, (1995).
\bibitem{Angilella} G.G.N. Angilella, R. Pucci and F. Siringo,  Phys. Rev.
{\bf B54}, 15471 (1996).
\bibitem{de Gennes} P.G.de Gennes, "Superconductivity of Metals and Alloys",
W.A.Benjamin, New York,1966.
\bibitem{Legget} A. J. Legget, Rev. Mod. Phys. {\bf 47}, 331, (1975).
\bibitem{Hng1} H. Gosh, Europhys. Lett. {\bf 43}, 707 (1998). 
\bibitem{Allgeier} C. Allgeier and J. S. Schilling, Physica {\bf C168},
499 (1990).
\bibitem{Marshall} D. S. Marshall, D. S. Dessau, A. G. Loeser, C-H Park,
A. Y. Matsura, J. N. Eckstein, I Bozovic, P. Fournier, A. Kapitulnik,
W.E Spicer and Z. X. Shen, Phys. Rev. Lett. {\bf 76}, 4841, (1996).
\bibitem{Loeser} A. G. Loeser, Z.X. Shen, D. S. Dessau, D. S. Marshall,
C. H. Park, P. Fournier and A. Kapitulnik, Science {\bf 273}, 325 (1996).
\bibitem{Ding2} H.Ding  T. Yokoya, J. C. Campuzano, T. Takahashi,
M. Randeria, M. R. Norman,
T. Mochiko, K. Kadowaki and J. Giapintizakis, Nature {\bf 382}, 51, (1996).
\bibitem{Norman} M. R. Norman,  H. Ding, M. Randeria, J. C. 
Campuzano, T. Yokoya, T. Takeuchi, T. Takahashi, T. Mochiko, K. Kadowaki,
P. Guptasarma and D. G. Hinks,  Nature {\bf 392}, 157, (1998).
\bibitem{Emery} V. J. Emery and S. A. Kivelson, Nature {\bf 374}, 434, (1995).
\bibitem{Coleman} P. Coleman  Nature {\bf 392}, 134, (1998).
\bibitem{Tallon} J. L. Tallon, C. Bernhard, H. Shaked, R.L Hitterman
and J. Jorgensen, Phys. Rev. {\bf B51}, 12911, (1995).

\end{references}
\end{document}